\documentclass[aps,pra,twocolumn,amsmath,amssymb]{revtex4-1}
\usepackage{color}
\usepackage{graphicx}
\usepackage{hyperref}
\usepackage{txfonts}
\usepackage{xcolor}

\newcommand{\mbp}[1]{\textcolor{black}{#1}}
\makeatother

\begin{document}

\title{
Enhancing the robustness of dynamical decoupling sequences with correlated random phases
}

\author{Zhen-Yu Wang$^{1,2}$, Jorge Casanova$^{3,4}$}
\author{Martin B. Plenio$^{5}$}
\affiliation{1. Guangdong Provincial Key Laboratory of Quantum Engineering and Quantum Materials,  School of Physics and Telecommunication Engineering, South China Normal University, Guangzhou 510006, China}
\affiliation{2. Frontier Research Institute for Physics, South China Normal University, Guangzhou 510006, China}
\affiliation{3. Department of Physical Chemistry, University of the Basque Country UPV/EHU, Apartado 644, 48080 Bilbao, Spain}
\affiliation{4. IKERBASQUE, Basque Foundation for Science, Maria Diaz de Haro 3, 48013, Bilbao, Spain}
\affiliation{5. Institut f\"ur Theoretische Physik und IQST, Albert-Einstein-Allee 11, Universit\"at Ulm, D-89069 Ulm, Germany}

\begin{abstract}
We show that \mbp{the addition of} correlated phases to the recently developed method of randomized dynamical
decoupling pulse sequences [Physical Review Letters {\bf 122}, 200403 (2019)] \mbp{can} improve its performance
in quantum sensing. \mbp{In particular, by correlating the relative phases of basic pulse units in dynamical
decoupling sequences, we are able to improve the suppression of the signal distortion due to $\pi$ pulse
imperfections and spurious responses due to finite-width $\pi$ pulses. This enhances selectivity of quantum
sensors such as those based on NV centers in diamond.}
\end{abstract}

\maketitle

\section{Introduction}
Dynamical decoupling (DD) techniques~\cite{viola1998dynamical} have important applications in quantum information
\cite{yang2010, suter2016RMP}, \mbp{quantum simulation \cite{Cai2013}} and quantum sensing~\cite{rondin2014magnetometry, wu2016diamond, suter2016single, degen2017RMP}. In particular, a sequence of DD $\pi$ pulses \mbp{is able to adjust
the resonance frequency of a qubit in a controlled manner} by periodically flipping its quantum state. In this manner,
the qubit \mbp{can be detuned from resonance} with respect to the frequencies of its surrounding noise, which results
in an extended coherence time. \mbp{At the same time,} when a DD sequence imprints a qubit flipping rate that matches
the frequency of a certain electromagnetic signal, the internal quantum state of the qubit gets modified leading \mbp{to} 
quantum detection. For example, under DD control the nitrogen-vacancy (NV) center qubit~\cite{doherty2013} in diamond has 
already demonstrated long coherence times~\cite{{ryan2010robust, deLange2010universal, BarGill2013, abobeih2018one}}, and 
an excellent sensitivity to AC magnetic fields~\cite{deLange2011single}. This makes  NV centers under DD control highly 
promising probes to detect, identify, and control nearby single nuclear spins~\cite{taminiau2012detection, kolkowitz2012sensing, zhao2012sensing, Muller2014, casanova2016noise, wang2017delayed, haase2018soft, lang2019PRL, bradley2019a} and spin clusters~\cite{zhao2011atomic, shi2014sensing, wang2016positioning,  abobeih2019atomic}.

One of the major factors limiting the performance of DD techniques is the unavoidable presence of errors in the applied control, 
which includes detuning and amplitude deviations on the applied  pulses. The effects of these errors can be partially compensated 
by robust DD sequences that make use of different pulse phases~\cite{gullion1990new,ryan2010robust,casanova2015robust,genov2017arbitrarily}. 
In addition, even if control pulses were ideally displayed (i.e. in absence of detuning and amplitude errors) the finite-width 
character of each pulse will introduce a spurious harmonic response~\cite{loretz2015spurious,haase2016pulse, lang2017enhanced, shu2017unambiguous}. \mbp{Importantly,} this spurious response \mbp{accumulates coherently} when a basic {\it DD pulse unit} is 
repeated $M>1$ times. At this point, it is important to remark that repeating several DD pulse units is the standard manner to 
achieve longer detection times. The accumulated spurious response can lead to a false identification of certain nuclear spins, 
e.g. the presence of $^{13}$C nuclei in a sample can be interpreted as the existence of $^{1}$H nuclei \mbp{as the ratio of 
the magnetic moments is almost precisely an integer}. All this has a negative impact on the reliability of general DD methods. 
In this respect, a recent theoretical and experimental study shows that the application of random global phases to each basic 
DD pulse unit of DD sequences suppress, up to some extent, the spurious harmonic response while enhances the sequence robustness 
against control errors~\cite{wang2019randomization}.

\begin{figure}
\center
\includegraphics[width=1.0\columnwidth]{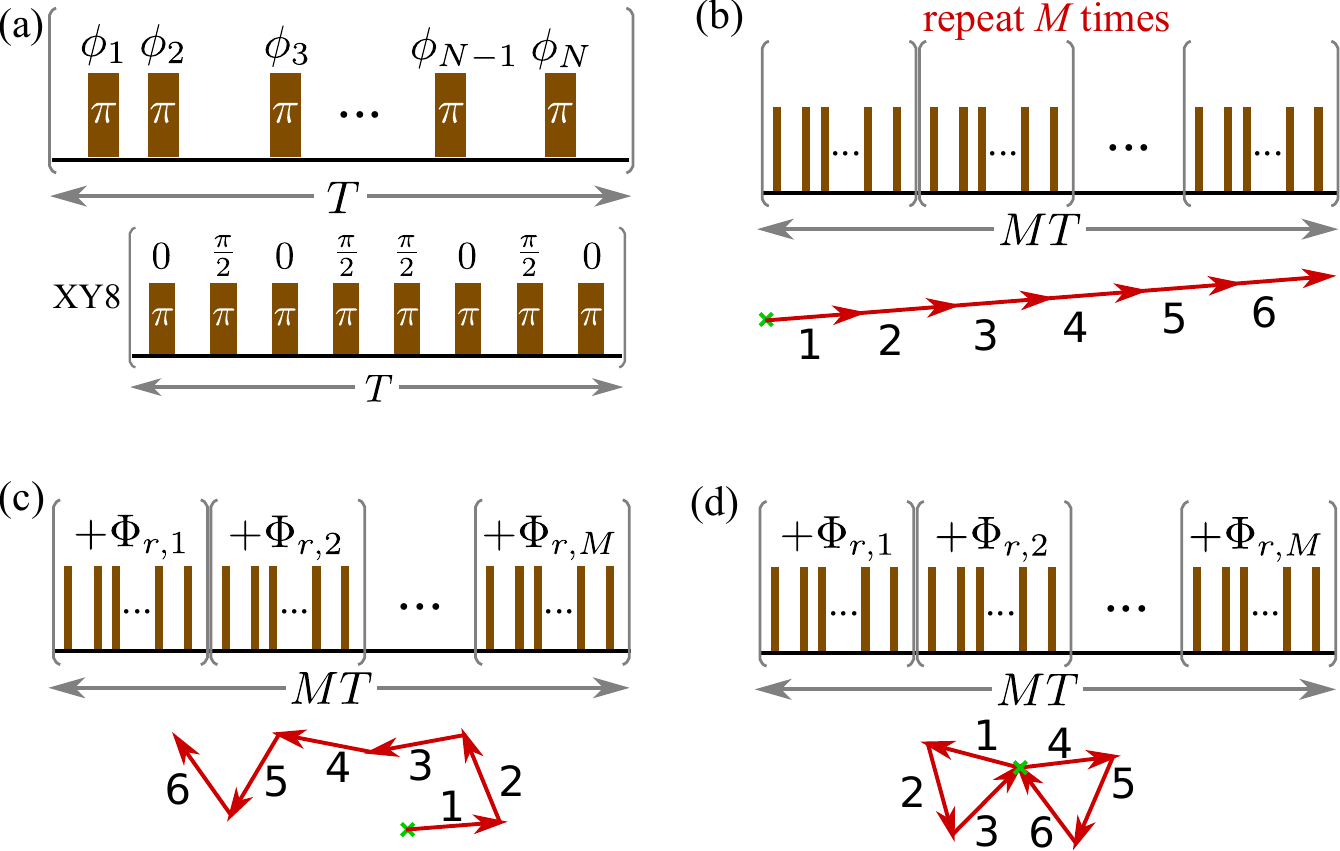}
\caption{Repetition of a basic DD pulse unit. (a) A basic unit of DD pulse sequence, which is defined by the positions and phases of the $\pi$ pulses. The lower panel is the example of an XY8 sequence.
(b) The standard protocol to construct a longer DD sequence is to repeat the same basic pulse unit illustrated in (a) $M$ times. Because the error contribution has the same phase factor for each DD pulse unit, the errors coherently add up (see the lower panel for the case of $M=6$).
(c) The randomisation protocol shifts all the pulses within each unit by a common, independent, random phase $\Phi_{r,m}$. Because of the random phases, the error terms of the basic DD pulse unit add up incoherently, suppressing the growing of error contribution (see the lower panel for an example).
(d) The correlated randomisation protocol imposes constraint on the phases on the random phases $\Phi_{r,m}$, such that the sum of their random phase factors vanishes. The lower panel illustrates an example that the sum of random phase factors of every three successive DD pulse units is zero.
}\label{Fig1}
\end{figure}

In this work, we introduce correlated phases to further enhance the performance of DD sequences for quantum sensing purposes. In particular, instead of using $M$ uncorrelated random phases (one for each of the $M$ basic DD pulse units present in a DD sequence) we impose constraints over these $M$ phases. This leads to a superior suppression of accumulated errors and spurious responses even when  the number $M$ of basic DD pulse units is small.

\section{Results}
\subsection{Effect of control imperfection}
Let us consider the periodic repetition of a basic pulse unit with a time duration $T$ and containing a number $N$ of $\pi$ pulses [see Fig. 1(a)]. In the case of ideal pulse sequences, each $\pi$ pulse is instantaneous and rotates the qubit by an angle $\pi$ along an axis in the $x-y$ plane. In realistic situations, however, the control Hamiltonian for the $\pi$ pulses in the rotating frame reads
\begin{equation}
\hat{H}_{c}=\frac{1}{2}\Omega\left[\hat{\sigma}_{x}\cos\phi+\hat{\sigma}_{y}\cos\phi\right]+\frac{1}{2}\Delta\hat{\sigma}_{z}, \label{Eq:Hc}
\end{equation}
where $\hat{\sigma}_{\alpha}$ ($\alpha=x,y,z$) is a Pauli matrix,  $\Omega$ is the Rabi frequency, and $\phi$ is the pulse phase of the control. Here the frequency detuning $\Delta$ introduces rotation axis and rotation angle errors. 	In addition, amplitude fluctuations on the control field change the value of the Rabi frequency $\Omega$, thus  further alters the rotation angle of the $\pi$ pulse.

For the sake of simplicity in the presentation, we do not consider the environment (e.g., a nuclear spin bath) of the qubit, and directly show the effect of pulse errors. However, the presence of nuclear spins will be taken into account in our numerical simulations of quantum sensing. Using Eq.~\eqref{Eq:Hc}, the evolution matrix of a single $\pi$ pulse has the general form ~\cite{genov2017arbitrarily,wang2019randomization}
\begin{equation}
\hat{U}_{\pi}(\phi)=\left(\begin{array}{cc}
e^{-i\alpha}\sin\epsilon & ie^{-i(\beta+\phi)}\cos\epsilon\\
ie^{i(\beta+\phi)}\cos\epsilon & e^{i\alpha}\sin\epsilon
\end{array}\right),
\end{equation}
where the real numbers $\alpha$, $\beta$, and $\epsilon$ depend on the explicit realization of the $\pi$ pulse but are independent of the pulse phase $\phi$.
We assume that each pulse has the same static errors, that is, $\alpha$, $\beta$, and $\epsilon$ are the same for all the $\pi$ pulses. When $\epsilon=0$ the pulse corresponds to a perfect $\pi$ pulse up to a drift $\beta$ (which can be caused by the detuning error $\Delta$) on the pulse phase. The pulse phase $\phi$ is fully tunable by changing the phase of the applied control field, and we assume that the phases of the pulses are applied in order with the values $\phi_1$, $\phi_2$, $\ldots$, $\phi_N$.

We consider widely-used basic DD $\pi$ pulse units leading to the identity operation on the qubit after their application.  Typical examples of these basic DD $\pi$ pulse units are the $\pi$ pulse arrangements belonging to the XY family~\cite{gullion1990new}, the YY8 sequence~\cite{shu2017unambiguous}, and the Carr-Purcell sequence~\cite{carr1954effects}, which contain \mbp{an} even number of $\pi$ pulses. To the first order of $\epsilon$, the evolution matrix of a DD pulse unit reads~\cite{genov2017arbitrarily,wang2019randomization}
\begin{equation}
\hat{U}_{{\rm unit}}=\left(\begin{array}{cc}
1 & iC\epsilon\\
iC^{*}\epsilon & 1
\end{array}\right)+O(\epsilon^{2}), \label{eq:UunitEven}
\end{equation}
where $C$ is a complex number that depends on the structure of the employed DD pulse unit. One example of a DD pulse unit (i.e. the widely used XY8 sequence) can be found in the lower panel of Fig.~\ref{Fig1}(a). In addition, \mbp{we note} that if one introduce a global phase shift $\Phi$ to the phases of all pulses the constant $C$ changes as $C\rightarrow C e^{-i \Phi}$.

\subsubsection{Standard protocol}
In the standard protocol where the basic pulse unit is repeated $M$ times as $\hat{U}  = (\hat{U}_{\rm unit})^M$, the control errors accumulate coherently. Using Eq.~\eqref{eq:UunitEven}, we obtain the evolution matrix of the whole sequence, i.e. of $\hat{U}$, this is:
\begin{equation}
\hat{U} = \left(\begin{array}{cc}
1 & i M C \epsilon \\
i M C^{*} \epsilon & 1
\end{array}\right)+O(\epsilon^{2}), \label{eq:Ustd}
\end{equation}
In this equation~\eqref{eq:Ustd} one can observe that the error $MC\epsilon$ scales linearly with $M$. This is illustrated in Fig.~\ref{Fig1}(b).

\subsubsection{Randomisation protocol}
In the randomisation protocol~\cite{wang2019randomization}, a random global phase $\Phi_{r,m}$ is imposed to all $\pi$ pulses of each $m$th  basic DD pulse unit. Now, by using Eq.~\eqref{eq:UunitEven}, the evolution matrix of the whole sequence reads~\cite{wang2019randomization}
\begin{align}
\hat{U}_{M} & =\hat{U}_{{\rm unit}}(\Phi_{r,M})\cdots \hat{U}_{{\rm unit}}(\Phi_{r,2})\hat{U}_{{\rm unit}}(\Phi_{r,1})\\
 & =\left(\begin{array}{cc}
1 & i Z_{r,M}MC\epsilon\\
iZ_{r,M}^{*}MC^{*}\epsilon & 1
\end{array}\right)+O(\epsilon^{2}), \label{eq:Urand}
\end{align}
where $Z_{r,M}=\frac{1}{M}\sum_{m=1}^{M}\exp(-i\Phi_{r,m})$, with  $\{\Phi_{r,m}\}$ being a set of phases. Due to the random value that  each phase $\Phi_{r,m}$ takes, the quantity $Z_{r,M}$ becomes a (normalised) 2D random walk with $\langle|Z_{r,M}|^2\rangle=1/M \leq 1$~\cite{wang2019randomization}. This suppresses the effect of control errors [see Fig.~\ref{Fig1}(c)].

\begin{figure}
\center
\includegraphics[width=1.0\columnwidth]{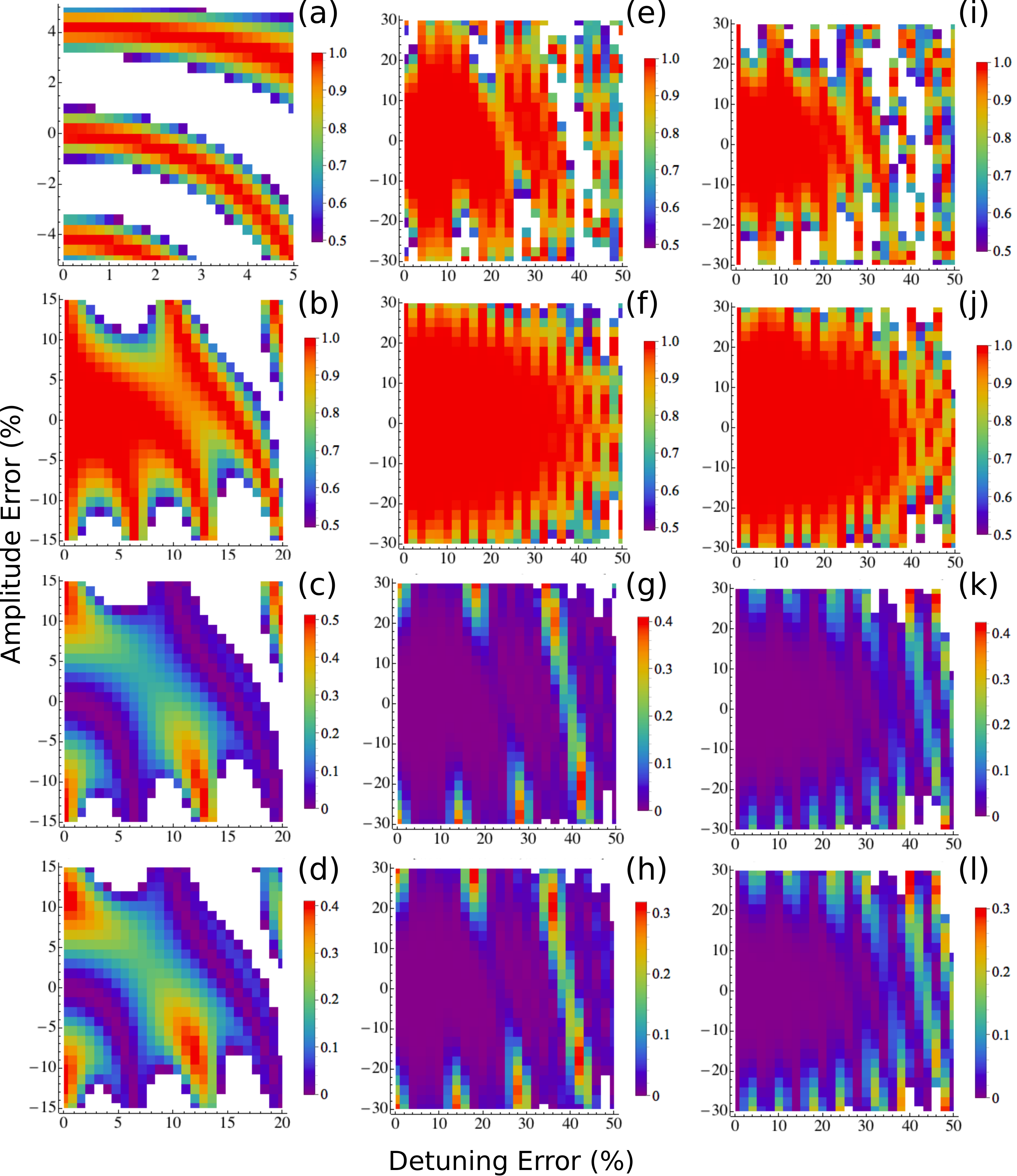}
\caption{Robustness of the protocols for a small $M$.
The fidelity (with respect to the identity) of \mbp{Carr-Purcell} sequences as a function of detuning and amplitude (Rabi frequency) errors for standard protocol (a) and correlated randomization protocol with the elimination size $G=2$ (b).  (c) and (d),  increased values of the fidelity when one changes the randomisation protocol in Ref.~\cite{wang2019randomization} to our correlated randomization protocol for elimination sizes $G=2$ and $G=3$, respectively.  (e)-(h) [(i)-(l)], as (a)-(d), but for the XY8 [YY8~\cite{shu2017unambiguous}] sequences.
In all figures, the regions in white have values out of the ranges shown in the color bars.
All the sequences consists of 48 $\pi$ pulses (that is, $M=6$ for XY8 and YY8 sequences). The time duration of each $\pi$ pulse is 15 ns, and the inter pulse spacing is 200 ns.
}
\label{FigRob48}
\end{figure}

\begin{figure}
\center
\includegraphics[width=1.0\columnwidth]{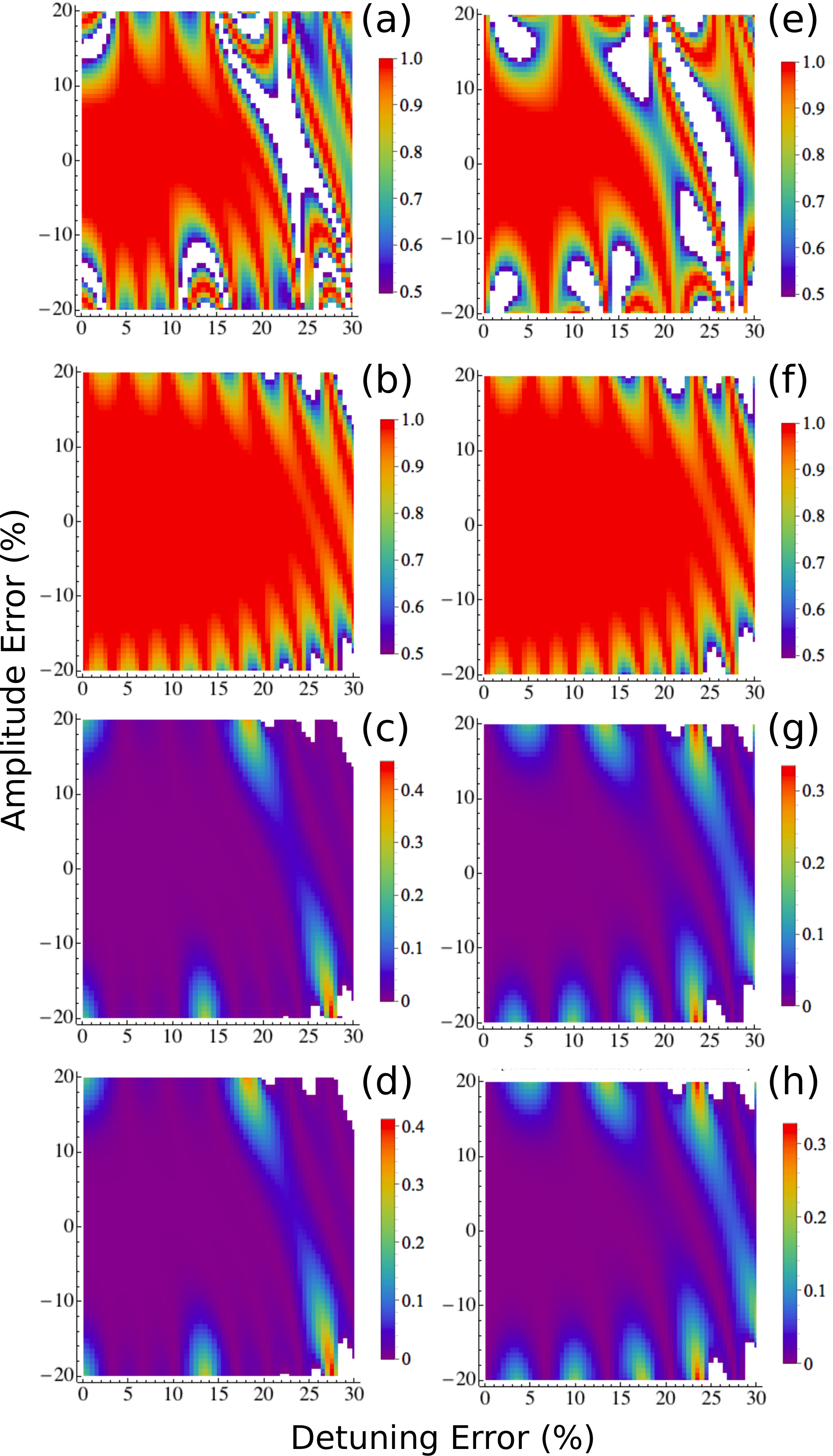}
\caption{Robustness of the XY8 and YY8~\cite{shu2017unambiguous} protocols for $M=24$.
(a)-(d) Results for XY8 sequences. (a) and (b) show the fidelity of XY8 sequences as a function of detuning and amplitude (Rabi frequency) errors for standard protocol (a) and correlated randomization protocol (b), respectively.  (c) and (d) are the fidelity enhancement over the randomization protocol by using the correlated randomization protocol for elimination sizes $G=2$ and $G=3$, respectively.  (e)-(h), as (a)-(d), but for YY8 sequences. The control parameters are the same as those used in Fig.~\ref{FigRob48} but with a larger $M=24$. In all figures, the regions in white have values out of the ranges shown in the color bars.}
\label{FigRobQ}
\end{figure}

\begin{figure}
\center
\includegraphics[width=0.95\columnwidth]{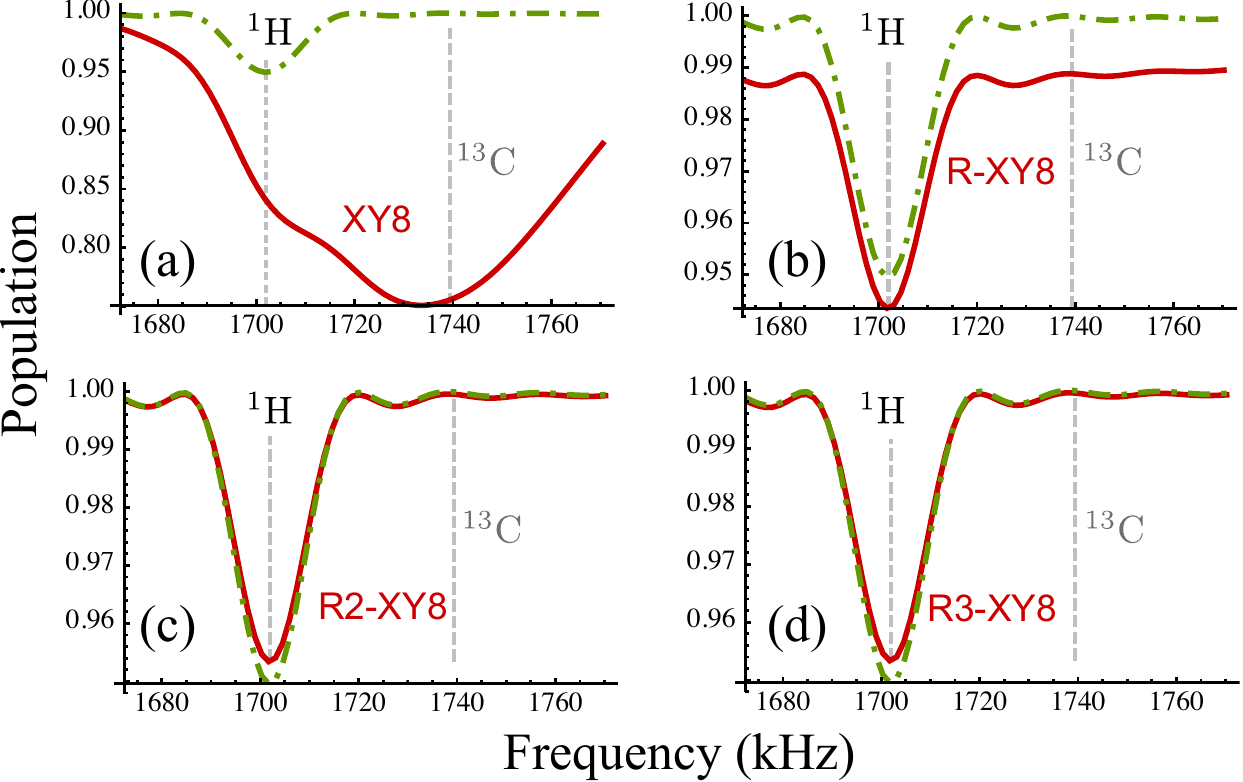}
\caption{Quantum spectroscopy with DD.
(a) Simulated averaged population signal (red solid line)
as a function of the DD frequency [$1/(2\tau)$ for pulse spacing $\tau$] for the standard XY8 protocol with a total number of $200$ $\pi$ pulses. The DD $\pi$ pulses have a non-zero time duration of 100 ns.  The frequency detuning and amplitude errors of the $\pi$ pulses have a static value of 10\% of the ideal Rabi frequency. The $^{1}{\rm H}$ spin to be sensed is coupled to the NV center via the hyperfine-field components~\cite{casanova2015robust} $(A_{\perp},A_{\parallel})=2\pi\times(2,4)$ kHz. A  $^{13}{\rm C}$ spin representing a noise source is coupled to the NV center via the hyperfine-field components~$(A_{\perp},A_{\parallel})=2\pi\times(10,200)$ kHz. The presence of $^{13}{\rm C}$ spin and imperfect control perturbs sensing signal and generates a spurious peak around ~1740 kHz (compare it with the green dash-dotted line obtained by an ideal error-free DD sequence).  (b) The use of randomization protocol suppresses the effect of pulse imperfection. The effect of errors is further suppressed in (c) and (d) by the use of correlated random phases elimination sizes $G=2$ and $G=3$, respectively. A magnetic field 400~G is used in the simulation.}
\label{FigS}
\end{figure}

\subsubsection{Correlated randomisation protocol}
Now, we impose a constraint on the random phases $\Phi_{r,m}$, such that $Z_{r,M}=\frac{1}{M}\sum_{m=1}^{M}\exp(-i\Phi_{r,m})=0$. In this manner, the effect of control errors \mbp{will} be suppressed more efficiently \mbp{as compared to a fully random scheme}, and irrespective \mbp{of} the value of $M$. Note that, in the randomization protocol in Ref.~\cite{wang2019randomization} a larger $M$ provides a better improvement as is demonstrated in the previous section with the expression $\langle|Z_{r,M}|^2\rangle=1/M \leq 1$. To cancel the effect of slowly fluctuating errors more efficiently, we  choose a smaller number $1<G\leq M$ of subsequent random phases such that $z_{r,G}=\sum_{j=1}^{G}\exp(-i\Phi_{r,k+j})=0$ for some integer $k$. Our method also improves the performance of the method over static errors since, by developing $\hat{U}_{M}$ in the perturbative parameter $\epsilon$, one can see that: Sequences with a low number $G$ better cancel high-order-error terms in $\epsilon$.

An example of a possible target sequence is in Fig.~\ref{Fig1}(d). Here we impose a constraint for every three subsequent random phases (i.e. $G=3$) such that the sum of their phase factors vanishes.  Note that it is not necessary to choose  the same value of $G$ for all the subsequent random phases in a single DD sequence. But for simplicity, we will use one fixed $G$ for each DD sequence in our simulations and call the value of $G$ as the {\em elimination size}.

\subsection{Comparison of different protocol performances}

In Fig.~\ref{FigRob48}, we compare the robustness of different protocols against control imperfections by numerically simulating the sequence fidelity. In particular, the fidelity is defined as the survival probability of the quantum sensor \mbp{in the initial state} (note this is initialised in an eigenstate of $\sigma_x$) after the application of the whole sequence. We use this definition because, in the absence of an external signal to detect, the quantum state of the sensor should remain unaffected after the application of the protocol. In addition, we clarify that white regions in these figures have lower fidelities (out of the range of each plot) such that their values are not shown for clarity.

An inspection to the panels in Fig.~\ref{FigRob48} shows that: Even with a small number of repetitions $M=6$ for the DD sequences, our method using correlated random phases presents an improved fidelity over the standard protocol and the randomisation protocol in Ref.~\cite{wang2019randomization}. For larger $M$ the fidelities of the randomisation protocol and our correlated random phases protocol are expected to be similar. However, still for $M=24$ our protocol is better than the randomisation protocol (see Fig.~\ref{FigRobQ}). We have also observed that our protocol using correlated random phases performs slightly better when one uses a small elimination size $G$ which is consistent with our theory [compare the results of $G=2$ and $3$ in Figs.~(\ref{FigRob48}) and (\ref{FigRobQ})] .

In Fig.~\ref{FigS}, we simulate the results of quantum sensing with XY8 sequences. In these simulations, we considered an NV quantum sensor in diamond. The target is to sense a proton ($^{1}{\rm H}$) spin outside of the diamond sample. Due to limited control power, the Rabi frequency $\Omega$ of control pulses has a finite value and, consequently, the $\pi$ pulses are not instantaneous and have a non-zero time duration. In addition, in our numerical simulations we also consider static errors on the control (see caption).  The non-zero pulse duration can leads to spurious resonances~\cite{loretz2015spurious} of sensing signal at the presence of other nuclear spins (e.g., a $^{13}{\rm C}$ spin in our simulation) [see Fig.~\ref{FigS}(a) for a simulation]. This signal error can be suppressed by the use of randomisation protocol [Figs.~\ref{FigS}(b)].  As shown in Figs.~\ref{FigS}(c), (d), the enhancement of the control robustness by using the correlated randomisation protocol can significantly further improve the signal fidelity in quantum sensing.

\section{Conclusion}
We showed that the robustness of randomisation DD protocol can be further improved if the random phases of the DD pulse units are chosen such that their phase factors have an average of zero. This reduces harmful effects due to  amplitude fluctuations, and frequency detuning of the DD control. In addition,
we have demonstrated that our correlated randomisation DD protocol provides better signals in DD based quantum sensing, even at the presence of nuclear spins that otherwise generate spurious peaks.

\emph{Acknowledgements.--}
M.~B.~P. acknowledges support by the ERC Synergy grant BioQ (Grant No. 319130), the EU project HYPERDIAMOND and AsteriQs, the QuantERA project NanoSpin, the BMBF project DiaPol, the state of Baden-W{\"u}rttemberg through bwHPC, and the German Research Foundation (DFG) through Grant No. INST 40/467-1 FUGG. J.~C. acknowledges financial support from Spanish Government via PGC2018-095113-B-I00 (MCIU/AEI/FEDER, UE), the  UPV/EHU grant EHUrOPE, the Basque Government via IT986-16, as well as from QMiCS (820505) and OpenSuperQ (820363) of the EU Flagship on Quantum Technologies, and the EU FET Open Grant Quromorphic.

\end{document}